\documentclass[twocolumn]{pasj00}
\draft
\SetRunningHead{T. Hayashi et al.}{Suzaku Observartion of V1223 Sgr}

\usepackage{times}
\usepackage{color}
\usepackage{rotating}

\newcommand{{\vIZ}}{V1223~Sgr}
\begin{document}

\title{Suzaku Observation of the Intermediate Polar V1223 Sagittarii}
\author{Takayuki \textsc{HAYASHI}$^{1,2}$, Manabu \textsc{ISHIDA}$^{1,2}$, 
  Yukikatsu \textsc{TERADA}$^3$, Aya \textsc{BAMBA}$^{4, 1, 5}$,\\
  and Takeshi SHIONOME$^{1,2}$
  }
\affil{
  $^1$  Institute of Space and Astronautical Science(ISAS)~/JAXA,
  3-1-1 Yoshinodai, Chuo-ku, Sagamihara, Kanagawa 252-5210\\
  thayashi@astro.isas.jaxa.jp\\
  $^2$  Department of Physics, Tokyo Metropolitan University,
  1-1 Minami-Osawa, Hachioji, Tokyo, 192-0397\\
  $^3$Department of Physics, Science, Saitama University, Sakura, 
  Saitama 338-8570\\
  $^4$School of Cosmic Physics, Dublin Institute for Advanced Studies,
  31 Fitzwilliam Place, Dublin 2, Ireland
  $^5$Department of Physics and Mathematics, College of Science and Engineering,
  Aoyama Gakuin University 5-10-1 Fuchinobe, Chuo-ku, Sagamihara,
Kanagawa, 252-5258, Japan

}

\maketitle

\begin{abstract}
 We report on the Suzaku observation of the intermediate polar V1223
 Sagittarii. Using a multi-temperature plasma emission model with its
 reflection from a cold matter, we obtained the shock temperature 
 to be 
 $37.9^{+5.1}_{-4.6}$~keV. This constrains the
 mass and the radius of the white dwarf (WD) in the ranges
$0.82^{+0.05}_{-0.06} \MO$ and $(6.9\pm 0.4) \times 10^8$~cm, 
 respectively, with the aid of a WD mass-radius relation. The solid
 angle of the reflector viewed from the post-shock plasma was measured
 to be
$\Omega/2\pi = 0.91\pm 0.26$.
 A fluorescent iron K$\alpha$ emission line
 is detected, whose central energy is discovered to be modulated with
 the WD rotation for the first time in magnetic-CVs.  Detailed spectral
 analysis indicates that the line comprises of a stable 6.4~keV
 component and a red-shifted component, the latter of which appears only
 around the rotational intensity-minimum phase. The equivalent width
 ($EW$) of the former stable component $\sim$80~eV together with the
 measured $\Omega$ indicates the major reflector is the WD surface, and
 the shock height is not more than 7\% of the WD radius. Comparing this
 limitation to the height predicted by the Aizu model (1973), we
 estimated the fractional area onto which the accretion occurs to be
 $<7\times 10^{-3}$ of the WD radius, which is the most
 severe constraint in non-eclipsing IPs.  The red-shifted iron line
 component, on the other hand, can be interpreted as emanating from the
 pre-shock accretion flow via fluorescence. Its $EW$
 ($28^{+44}_{-13}$~eV) and the central energy
 ($6.30_{-0.05}^{+0.07}$~keV) at the intensity-minimum
 phase are consistent with this interpretation.
\end{abstract}

\section{Introduction}
\label{sec:intro}

Cataclysmic variables (CVs) are binaries composed of a white dwarf (WD)
and a main sequence star or a red giant that fills its Roche lobe. Of
them, a group of systems harbouring a magnetized WD ($B < 10$~MG)
whose rotation is not synchronized with the secondary rotation is
referred to as an intermediate polar (IP). In the IPs, matter spilt over
the Roche lobe of the secondary star is funneled into the accretion
columns by the strong magnetic field within the Alfv$\acute{\rm e}$n
radius, and falls toward the WD surface nearly at the free fall speed.
Since the accreting matter becomes highly supersonic as it descends the
column, a strong standing shock is formed close to the WD surface, and
the matter is heated up to a temperature of
\begin{equation}
kT_{\rm S} \;=\; \frac38\frac{GM}{R}\mu m_{\rm H}
 \;=\; 16\,\left(\frac{M}{0.5M_\odot}\right)
 \,\left( \frac{R}{10^9\mbox{cm}}\right)^{-1}\;\;\mbox{[keV]}
\label{eq:shockT}
\end{equation}
\citep{1973PThPh..49..776H,1973PThPh..50..344A}. The high temperature
plasma thus formed is cooled via optically thin thermal plasma emission,
and finally settles onto the WD surface. Since the accretion is
inhomogeneous, the X-ray intensity from the IPs are modulated at the WD
rotational period.
The modulation is considered to be due mainly to photoelectric
absorption by the pre-shock accreting matter. As a result, the
modulation depth increases with decreasing X-ray energy, and the
pulse-intensity-minimum occurs when one of the accretion columns becomes
closest to the observer's line of sight \citep{1988MNRAS.231..549R}.

Since the maximum temperature of the plasma is proportional to the
gravitational potential of the WD as shown in Eq.~(\ref{eq:shockT}), the
WD mass can be estimated by evaluating the maximum temperature with an
X-ray continuum measurement. Although this approach has been tried so far
in a number of works
\citep{2004A&A...419..291B,2004A&A...426..253R,2005A&A...435..191S,2007ApJ...663.1277E,2009A&A...496..121B,2010A&A...520A..25Y},
it is not so straightforward. First, the temperature distribution
described above should be taken into account in the emission
model. Second, the observed X-ray emission is contaminated by the plasma
emission reflected off from 
the WD surface as well as the pre-shock accreting matter
\citep{1999ApJS..120..277E}, whose spectra is significantly different
from the direct spectrum. Third, the line-of-sight hydrogen column
density also has a distribution with a maximum column of $N_{\rm H}
\simeq 10^{23.5}$~cm$^{-2}$ \citep{1999ApJS..120..277E}.
All these effects should be carefully considered in evaluating the mass
from the X-ray continuum emission.
 
On the other hand, the intensity of the reflected continuum is useful in
evaluating the geometry of the hot plasma. In addition to the continuum,
an iron emission line at 6.4~keV emanates from the same reflector via
fluorescence. Its equivalent width ($EW$) is useful for evaluating
the solid angle of the reflector viewed from the plasma
\citep{Maxima,1991MNRAS.249..352G}. 
Since part of the 6.4~keV line is emitted from the pre-shock accreting
matter, its central energy modulation at the WD rotational period may be
detected from some IPs.

In order to measure the WD mass more accurately than ever and to
elucidate the geometry of the accretion column by means of a better
evaluation of the intensities of the reflected continuum and the 6.4~keV
iron line,
we have carried out the observation of the typical IP V1223 Sagittarii
(4U1849-31) with Suzaku \citep{2007PASJ...59S...1M}. High spectral
resolution of the X-ray Imaging Spectrometer (XIS) as well as high
sensitivity of the Hard X-ray Detector (HXD) above 10~keV are ideal
combination to carry out this investigation. {\vIZ} locates at $\ell
= 4^{\circ} \timeform{57'28.9''}, b = -14^{\circ} \timeform{21' 16.8''}$
which was identified by HEAO 1 \citep{1981ApJ...249L..21S}. {\vIZ}
shows a pulsation at the spin period of the WD of $P_{\rm spin} =
745.6$~s \citep{1985SSRv...40..143O}, and an intensity modulation at an
orbital period of $P_{\rm orb} = 3.37$~hr \citep{1987ApJ...323..672J}.
Additional intensity modulation at the beat period between them ($P_{\rm
beat} = 794.4$~s) is also detected \citep{1981ApJ...249L..21S}. The
distance and the orbital inclination are known to be $D = 527$~pc and $i
= 24^\circ$, respectively \citep{2004A&A...419..291B}.  The mass of the
WD evaluated from the previous works, on the other hand, distributes in
a wide range of 0.71--1.046~$\MO$ (see table~\ref{table:mass}).

This paper is organized as follows. In \S~\ref{sec:obs}, we introduce the
instruments onboard Suzaku and show the observation log and the procedure
of the data reduction. In \S~\ref{sec:ana}, details of our timing
and spectral analysis are presented. The high spectral resolution
and the wide-band coverage
enable us to determine the X-ray spectral parameters with unprecedented
accuracy. In \S~\ref{sec:dis}, we discuss the mass of the WD and the
geometry of the post-shock plasma in {\vIZ}, and the spin modulation
of the fluorescent iron K$\alpha$ line
In \S~\ref{sec:con}, we summarize our results and discussions.

\section{Observation and Data Reduction}\label{sec:obs}

\subsection{Observation}

\begin{longtable}{*{7}{c}}
  \caption{Suzaku observation log of V1223 Sagittarii.}\label{table:obs}
  \endhead
  \hline
  Sequence $\sharp$ & Observation date (UT) & Pointing & Detector$^{\ast}$ & Mode & Exposure$^{\dagger}$ & Intensity$^{\ddagger}$\\
  &&&&&(ksec)&(count s$^{-1}$)$^{\S}$\\\hline
  402002010 & 2007 Apr 13 11:31-14 22:36 & HXD nom. & XIS & Normal & 60 sec & 10.54$_{-0.02}^{+0.03}$\\
  &&& HXD-PIN & Normal & 46 sec & 0.57$\pm$0.01\\\hline
  \multicolumn{7}{l}{\footnotesize 
    $^{\ast}$XIS: XIS0,1,3 are combined.}\\
  \multicolumn{7}{l}{\footnotesize 
    $^{\dagger}$Exposure time after data screening.}\\
  \multicolumn{7}{l}{\footnotesize
    $^{\ddagger}$Intensity in the 0.1-11.5~keV for the XIS
    and in the 12-50~keV band for the HXD-PIN with 1$\sigma$ error.}\\
  \multicolumn{7}{l}{\footnotesize
    $^{\S}$ XIS0, 1, 3 are combined.}
  \endfoot
\end{longtable}

The Suzaku observation of {\vIZ} (Seq.\#402002010) was carried out
from 2007 April 13 11:31 (UT) to April 14th 22:36 (UT).
The observation log is summarized in table~\ref{table:obs}.
Suzaku is equipped with four modules of the XIS
\citep{2007PASJ...59S..23K}, which are designed as XIS0, XIS1, XIS2, and XIS3.
Of them, XIS1 adopts a back-illuminated (BI) CCD, while the other three
employ front-illuminated (FI) CCDs. They cover energy ranges of
0.2--10~keV and 0.4--10~keV, respectively. The energy resolution of them
is $\sim$150~eV (FI) and $\sim$160~eV (BI) in FWHM at 6 keV at the time
of this observation.\footnote{http://www.astro.isas.jaxa.jp/suzaku/process/caveats/caveats\_xrtxis06.html}
The energy calibration accuracy at the Mn-K${\alpha}$ line (5.895 keV)
is $\lesssim$5 eV.\footnote{http://www.astro.isas.jaxa.jp/suzaku/doc/suzakumemo/suzakumemo-2007-06.pdf}$^,$
\footnote{http://www.astro.isas.jaxa.jp/suzaku/analysis/xis/sci/} 
One of the modules XIS2 has been unusable since 2006 November, and hence we
use the other three XIS modules in this paper.

Each XIS module is located in the focal plane of an X-ray telescope
(XRT: \cite{2007PASJ...59S...9S}). The XRT adopts Wolter-I type
grazing-incident reflective optics consisting of tightly nested,
thin-foil conical mirror shells. The angular resolution ranges from
\timeform{1'.8} to \timeform{2'.3} in half-power diameter. The effective
area is 440 cm$^2$ at 1.5 keV and 250 cm$^2$ at 8 keV per XRT module.

The energy range above 10~keV is covered with the HXD, which is a
non-imaging, collimated detector
\citep{2007PASJ...59S..35T,2007PASJ...59S..53K}.
It is composed of the two detectors. One is the PIN detector which
adopts 2~mm-thick silicon PIN diodes, and is sensitive to an X-ray
energy in 10--70~keV.
the other is a GSO/BGO phoswitch counter which is sensitive in the
40--600~keV band.
The energy resolution is 3.0 keV (FWHM) for the PIN detector, and
7.6/$\sqrt[]{\mathstrut E}$\% (FWHM) for the GSO detector, where $E$ is
energy in MeV. Since we have detected no significant flux with the GSO,
we only use the PIN data in this paper.

Throughout the observation of {\vIZ}, the XIS was operated in the normal
5$\times$5 and 3$\times$3 editing modes during the data rate SH/H and
M/L, respectively. Spaced-row Charge Injection (SCI)
\citep{2008PASJ...60S...1N} is applied while no window/burst options is
used. The HXD PIN was operated with a bias voltage of 500 V for 8 out of
64 modules and 400 V for the others to suppress the rapid increase of
noise events, possibly caused by in-orbit radiation damage.

The observation was performed at the HXD nominal position in order to
collect more photons over 10 keV where the reflection continuum stands out.

\subsection{Data Reduction}\label{sec:reduction}

\begin{figure}[h!]
  \begin{center}
    \vspace{-5mm}
    \hspace{5mm}
\FigureFile(80mm,80mm){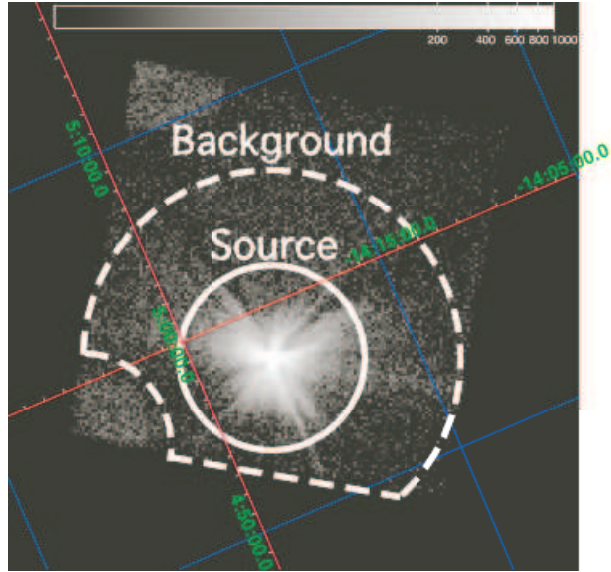}
  \end{center}
\caption{The image of the {\vIZ} with XIS0 in the galactic coordinate,
 and the integration regions of source and background photons. The gray
 scale represents the X-ray surface brightness in a logarythmic
 scale. The source region is a circle with a radius of \timeform{4'34}
 centered on the source, while the background region is an annulus with
 an outer radius of \timeform{9'} out of the source integration
 region. The bright regions at the upper and lower left corners are
 irradiated with the onboard $^{55}$Mn sources.}  \label{fig:image}
\end{figure}

We used the datasets produced by the Suzaku pipe-line processing version
2.0.6.13 with the calibration files of hxd20070710, xis20070731, and
xrt20070622 for the HXD, XIS, XRT, respectively. We analyzed the data
with the analysis software package HEASOFT version 6.3.1 and XSPEC
version 12.5.1 \citep{1996ASPC..101...17A}. For the XIS analysis, we
employed photons with the ASCA grades of 0, 2, 3, 4, 6 events. We
discarded the data while the satellite telemetry was saturated and the
telemetry data rate was L in which the telemetry usually
saturates. Furthermore we excluded the data taken while the earth
elevation angle is less than 5$^\circ$ (ELV$<$5), the day-earth
elevation angle is less than 20$^\circ$ (DYE\_ELV$<$20), and the
spacecraft passes in the South Atlantic Anomaly. As a result, the total
exposure time of the XIS is about 60~ks. We employed a circular region
with a radius of 250 pixels (\timeform{4'.34}) centered on the {\vIZ}
image as the integration region of the source photons, which includes
more than 96\% flux from the source. The background photons are
collected from an annulus between radii of 250 pixels and 500 pixels
centered on {\vIZ}. The source and the background regions are drawn on
the observed X-ray image in Fig.~\ref{fig:image}.
The background-subtracted mean intensity of the XIS is
$10.54^{+0.03}_{-0.02}$ counts sec$^{-1}$ in 0.1-11.5 keV with the three
XIS modules.

We employed nearly the same data selection criteria to the HXD as to the XIS.
We, however, did not apply the day-earth elevation criterion
(DYE\_ELV$<$20), but instead, we discarded the data while the cutoff
rigidity is less than 6~GeVc$^{-1}$. As a result, we obtained about 46 ksec
exposure time for the HXD PIN.

We adopted the ``tuned'' non-X-ray background (NXB) event file provided
by the HXD team, using METHOD = ''LCFITDT (bgd\_d)'' and version of
METHODV = ``2.0ver0804'' \citep{2009PASJ...61S..17F}, and synchronized
GTIs (good time intervals) of this with that of the source event
file. Eventually, averaged intensity in energy band of 12-50 keV after
the NXB subtraction is 0.57 $\pm$ 0.01 c~s$^{-1}$. In addition to this,
for spectrum analysis, we considered cosmic X-ray background (CXB).
Following results based on HEAO-1 observation
\citep{1987PhR...146..215B}, we assumed the CXB spectrum of
\begin{eqnarray} 
f_{{\rm CXB}(E)}
 = 9.0 \time 10^{-9}
 \times \left(\frac{E}{3 {\rm keV}}\right)^{-0.29}
 \times {\rm exp} \left( -\frac{E}{40 {\rm keV}}\right)
        {\rm erg \hspace{0.5 ex} cm^{-2} s^{-1} str^{-1} keV^{-1}}.
\label{eq:cxb}
\end{eqnarray} 
According to this equation, the intensity of the CXB is found to be
equal to $\sim$5 \% of the NXB in the 12--50~keV band. We created a CXB
spectrum file from Eq.~(\ref{eq:cxb}) by the {\tt fakeit} command 
in XSPEC using HXD PIN flat sky response file {\tt
ae\_hxd\_pinflat3\_20081029.rsp}, assuming that the uniform emission is
from a sky region of $2^\circ \times 2^\circ$
\footnote{http://heasarc.gsfc.nasa.gov/docs/suzaku/analysis/pin\_cxb.html}.

\section{Analysis and Results}\label{sec:ana}

\subsection{Timing Analysis}

\begin{figure}[!h]
  \begin{center}
    \FigureFile(80mm,50mm){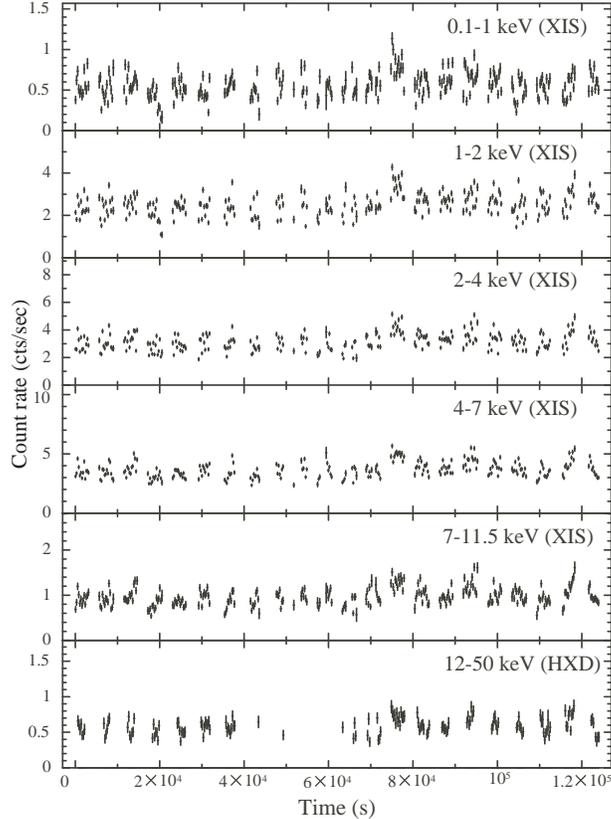}
  \end{center}
\vspace{-0.5cm}
  \caption{Energy-resolved light curves of {\vIZ} subtracted
 the background with the image region (Fig.\ref{fig:image})
 in the bands 0.1-1~keV, 1-2~keV, 2-4~keV, 4-7~keV, and
 7-11.5 keV with the XIS0, 1 and 3 being combined, and only NXB subtracted 12-50~keV from
 the HXD-PIN. The bin size is 256 sec. The vertical axis of each panel
 is scaled so that the maximum value is three times as large as the
 average counting rate. The origin of the time is MJD 54203.50068.}
 \label{fig:lc}
\end{figure}

\begin{figure}[h!]
  \begin{center}
    \FigureFile(80mm,50mm){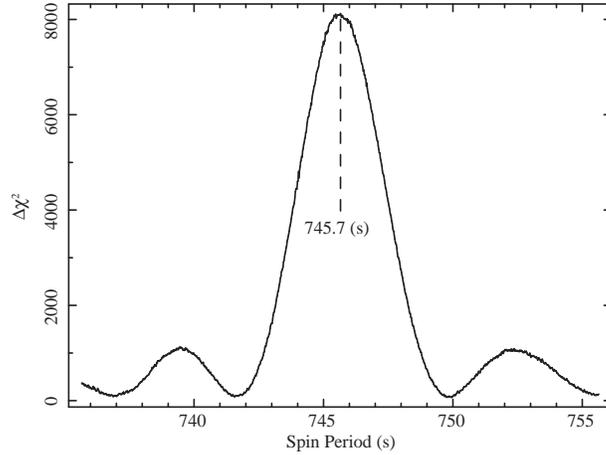}
  \end{center}
\vspace{-0.5cm}
 \caption{Periodogram calculated from background subtracted XIS light
 curve in the 0.5-11.5 keV energy band. The horizontal and vertical axes
 show trial periods and $\chi^2$, respectively. The $\chi^2$ values are
 evaluated with light curves with 31 bin/cycle. The trial period step is
 0.01~s. The maximum $\chi^2$ is obtained at the period 745.7~s.}
 \label{fig:pg}
\end{figure}

\begin{figure}[h!]
  \begin{center}
    \FigureFile(80mm,50mm){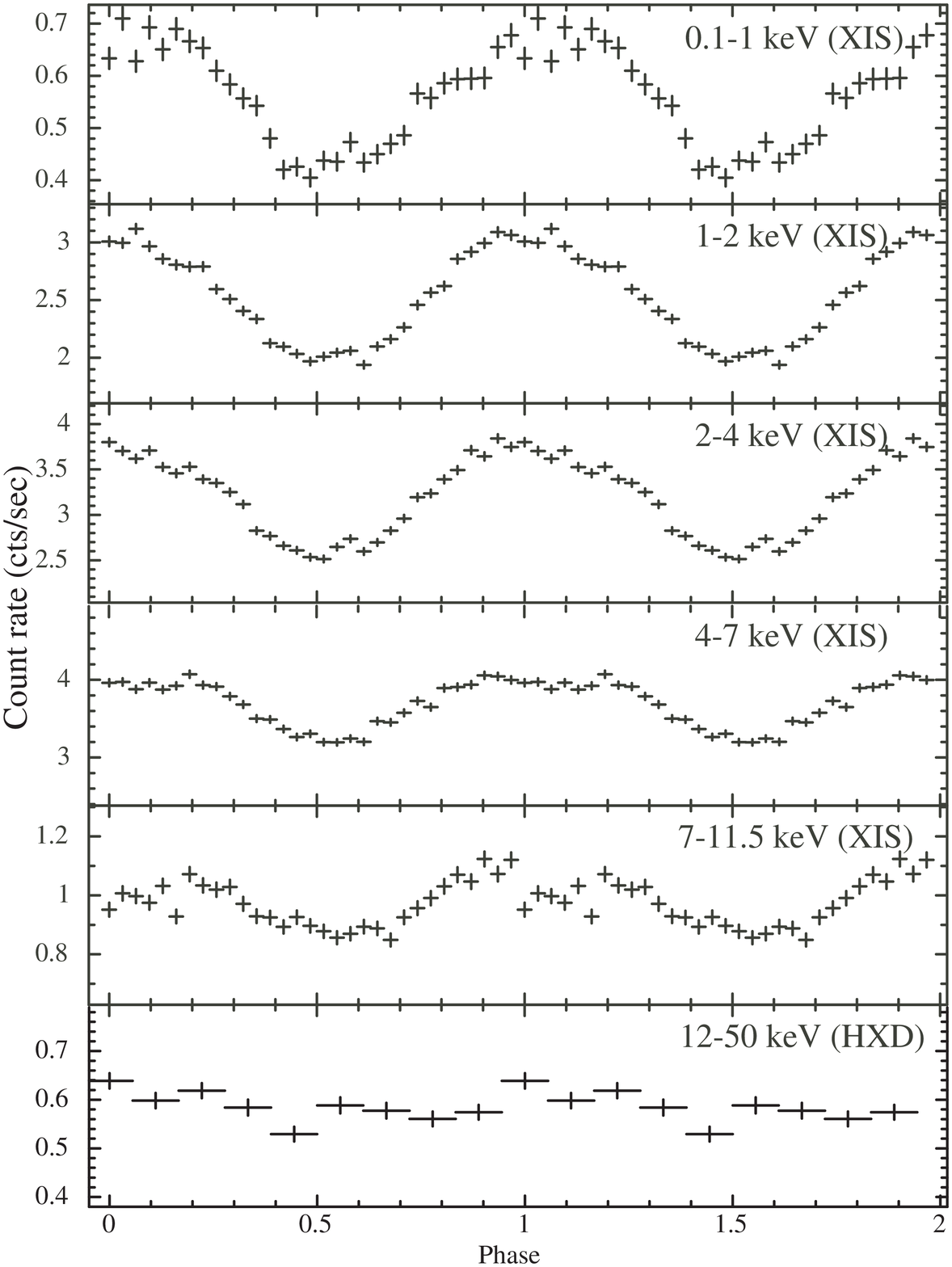}
  \end{center}
\vspace{-0.5cm}
  \caption{Folded light curves at the period of 745.7~s. They consist of
 31 bins and 9 bins per cycle for the XIS and the HXD-PIN,
 respectively. The vertical axis of each panel is scaled so that
 the minimum and the maximum values are 0.65-times and 1.35-times
 of the averaged counting rate, respectively.} \label{fig:flc}
\end{figure}

\begin{figure}[h!]
  \begin{center}
    \FigureFile(80mm,50mm){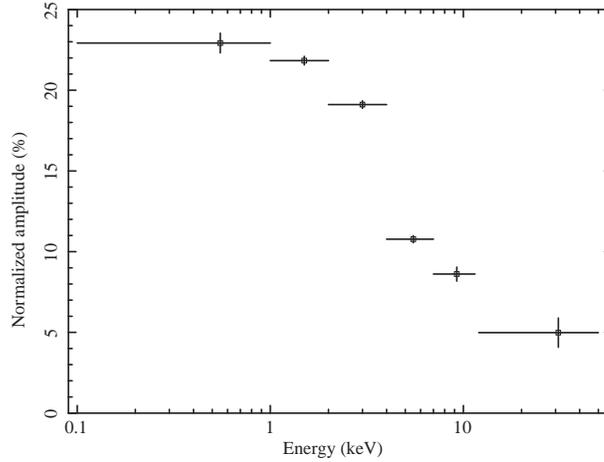}
  \end{center}
\vspace{-0.5cm}
  \caption{Fractional spin-modulation amplitude as a function of the X-ray
 energy.} \label{fig:amp}
\end{figure}

Fig.~\ref{fig:lc} shows the XIS light curves in the bands 0.1--1
keV, 1--2 keV, 2--4 keV, 4--7 keV, and 7--11.5 keV, where all the
available modules (XIS0, 1, and 3) are combined, and that of the HXD-PIN
in the 12--50 keV band.
Those of the XIS are after background subtraction with the image region
shown in Fig.~\ref{fig:image}, whereas only the NXB is subtracted in the
HXD PIN light curve.
Significant signals were detected in all the energy bands. There is an
intensity variation associated with the WD rotation (= 746~s) as well as
a long term one with a time scale of $\sim$$10^{4}$~s. In order to
evaluate the period of the WD spin, we first carried out an FFT analysis
with the barycentric corrected light curve of the XIS, and identified
a rough spin period. We then performed an epoch-folding analysis and
produced a periodogram, which is shown in Fig.~\ref{fig:pg}. 
As a results, we got the period of 745.7 $\pm$ 1.1 sec, which is
consistent with 745.6 sec \citep{1987ApJ...323..672J} measured with an
optical photometry.  
Fig.~\ref{fig:flc} shows light curves folded at the spin period thus
determined. 
The energy bands are common with those adopted in
Fig.~\ref{fig:lc}. The profile of the folded right curves are almost
sinusoidal in the energy band below 4 keV. In the energy band
above 4 keV, on the other hand, a flat profile appear at around the peak
phase of the modulation. As shown in Fig.~\ref{fig:amp}, the modulation
depth evaluated by fitting the light curves with constant plus sinusoid
decreases with increasing energy, which is consistent with the
previous results (\cite{1985SSRv...40..143O}; Taylor et al. 1996). 

\subsection{Average Spectrum}\label{sec:avespe}

\begin{figure}[h!]
  \begin{center}
    \FigureFile(80mm,50mm){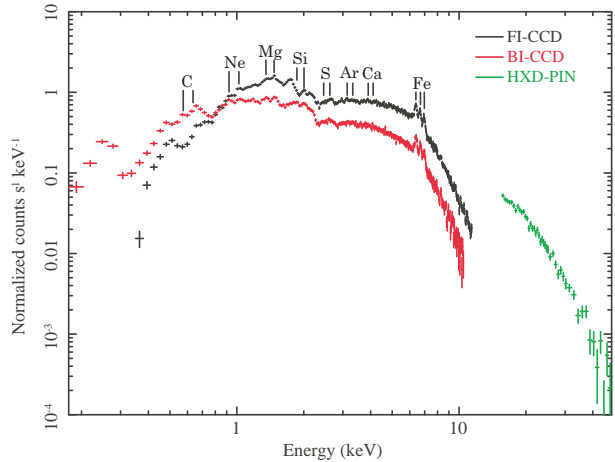}
  \end{center}
\vspace{-0.5cm}
  \caption{The averaged spectra of {\vIZ}. The spectrum of the BI-CCD
 (XIS1), the FI-CCDs (XIS0 + XIS3) and the HXD-PIN are shown in red,
 black, and green, respectively.} \label{fig:avespe}
\end{figure}

Fig.~\ref{fig:avespe} shows the averaged spectra of {\vIZ} after 
background subtraction according to \S~\ref{sec:reduction}.
The spectra from the two FI-CCD modules (XIS0 and XIS3) are co-added.
The source is detected up to 50 keV by the grace of the high sensitivity
of the HXD-PIN. In addition to many emission lines from highly ionized
metals in the post-shock plasma, a fluorescent iron K$\alpha$ line at
6.4 keV is also prominent, which indicates existence of reflecting cold
matter, such as the WD surface and the pre-shock accretion matter.

In order to constrain physical quantities, 
we have constructed the following model, and fit it
to the averaged spectra. We adopted the optically
thin thermal plasma emission model {\tt cevmkl}
\citep{1997MNRAS.288..649D} in XSPEC derived from the {\tt mekal} model
\citep{1985A&AS...62..197M,1986A&AS...65..511M,1995ApJ...438L.115L,1996uxsa.conf..411K}
to represent the temperature distribution in the post-shock plasma
(\S~\ref{sec:intro}). The {\tt cevmkl} model introduces the power-law
type differential emission measure as
\begin{eqnarray}
  d(EM) \propto \left(\frac{T}{T_{\rm max}} \right)^{\alpha} d({\log}T) 
  \propto \left(\frac{T}{T_{\rm max}}  \right)^{\alpha-1}dT,
\end{eqnarray}
where $T_{\rm max}$ is the maximum temperature of the plasma. 
Abundances of the major metals can be set separately
in the model. On the assumption that the post-shock
accretion column is isobaric \citep{1992Sci...258.1015F}, $\alpha$
becomes equal to 0.5 \citep{1994MNRAS.266..367I}. 
On the other hand, full theoretical analysis, which takes the pressure
gradient and the gravity effect fully into account, predicts $\alpha =
0.43$ \citep{2005A&A...435..191S,2005A&A...444..561F}.

Since we detected the fluorescent iron K$\alpha$ line at 6.4~keV, we
need to take into account a reflected continuum of the plasma emission.
To represent the reflection continuum,
we adopted the model {\tt reflect} \citep{1995MNRAS.273..837M} in XSPEC
which is a convolution-type model describing the reflectivity of a neutral
material.
The {\tt reflect} model considers a situation in which a point-like
emission source illuminates a thick slab with a solid angle of $\Omega$,
and an observer views the slab with an inclination angle of $i$. We can
set the iron abundance and that of the other elements separately.
In the spectral fit, we fixed $i$ to the orbital inclination angle
24$^\circ$ \citep{2004A&A...419..291B} {as an average
over the WD spin phases.
Since the {\tt reflect} model does not include a fluorescent iron line
due to its convolution-type nature, we added a Gaussian emission line at 6.4 keV.

In IPs, the spectra undergo strong photoelectric absorption
\citep{1989MNRAS.237..853N,1999ApJS..120..277E} which can not be
characterized by a single column density.
The modeling of the absorption affects the
characterization of the intrinsic spectral shape. It is, however, shown
that we can avoid this difficulty by using only a moderately high
energy band \citep{1999ApJS..120..277E} where the multi-column densitise can be
approximated by a single column density. In the analysis hereafter, we
use the energy band $>$5~keV for the XIS.
We adopted the {\tt phabs} model in XSPEC in representing the single
photoelectric absorption column density.

In the model, we refer to \citet{1989GeCoA..53..197A} as the solar
abundances of the metals. The abundances of iron and nickel of
the {\tt cevmkl} model and that of iron  of the {\tt reflect} model are
constrained to be the same.
In addition, the abundance except for iron of the {\tt reflect} model is
synchronized to that of oxygen of the {\tt cevmkl} model, because the surface of
the WD is probably covered with the accreted matter and oxygen 
modifies the reflected continuum, next to iron.  Since the other
elements do not affect the reflected continuum so severely, their
abundances are set equal the solar abundances.

The best-fit model overlaid on the spectra is shown in
Fig.~\ref{fig:avespe_fit}, and its parameters are summarized
in table~\ref{table:ave_para}.
\begin{figure}[h!]
  \begin{center}
    \FigureFile(80mm,50mm){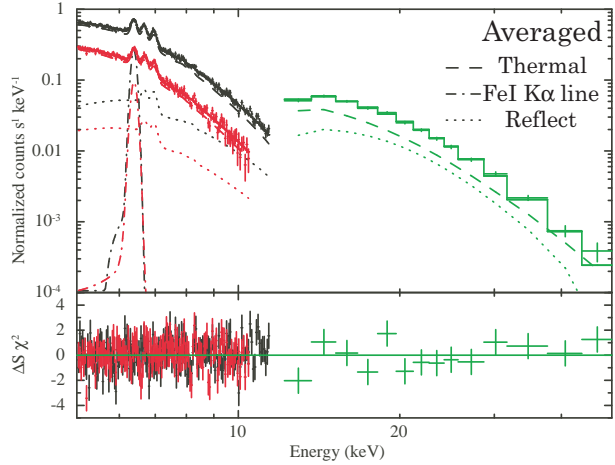}
  \end{center}
  \caption{The best-fit {\it
    phabs$\times$(reflect$\times$cevmkl+Gaussian)} model overlaid on the
    averaged spectra. Black, red and green colors are the same as in
    Fig.~\ref{fig:avespe}. Crosses, solid lines, dashed lines, dotted
    lines, and dotted-dashed lines represent the data points, the total
    model, the intrinsic thermal emission model ({\it cevmkl}), its
    reflection ({\it reflect}) model, and the fluorescent iron K$\alpha$
    line ({\it Gaussian}). The lower panel shows the fit residual in a
    unit of $\sigma$.} \label{fig:avespe_fit}
\end{figure} 
\begin{table}
  \caption{Best-fit parameters of a simultaneous fit to Suzaku averaged
    spectrum. The adopted model is {\it phabs}$\times${\it
    reflect}$\times$({\it cevmkl}+ {\it Gaussian}).$^{\ast}$
    \label{table:ave_para} }
  \begin{center}
    \vspace{0.2cm}
    \begin{tabular}{ccccc}\hline
      Parameter & \multicolumn{2}{c}{Value} \\\hline
      $N_{\rm H}$ ($\times10^{22}$ cm$^{-2}$) & $8.9_{-1.7}^{+1.8}$ & $9.1_{-1.6}^{+1.7}$\\
      $T_{\rm max}^\dagger$ (keV) & $33.5_{-6.1}^{+7.1}$ & $37.9_{-4.6}^{+5.1}$\\
      $\alpha^{\ddagger}$ & $0.58_{-0.17}^{+0.19}$ & 0.43 (fixed)\\
      $\Omega/(2\pi)^{\S}$ & $0.95_{-0.27}^{+0.40}$ & $0.91\pm 0.26$\\
      $i^{\parallel}$ (deg)&24 (fixed) & 24 (fixed)\\
      $Z_{\rm O}^{\sharp}$ & $1.07_{-0.39}^{+0.71}$ & $0.86_{-0.26}^{+0.40}$\\
      $Z_{\rm Fe}^{\sharp}$& $0.29\pm0.03$ & $0.28_{-0.02}^{+0.03}$\\
      FeI-K$\alpha$ line Energy (keV) & $6.391_{-0.005}^{+0.006}$ & $6.391_{-0.004}^{+0.006}$\\
      FeI-K$\alpha$ line sigma (eV) & 35 $_{-16}^{+13}$ & $33_{-16}^{+10}$\\
      FeI-K$\alpha$ line $EW$ (eV) & $99_{-11}^{+16}$ & $97_{-12}^{+13}$\\
      $\chi^2$ (d.o.f.) & 350.78 (313) & 351.90 (314)\\\hline
    \end{tabular}
\end{center}
      {\footnotesize 
        \hspace{1cm}
        All the uncertainties are at the 90\% confidence level.\\
        \hspace{1cm}
        $^{\dagger}$Maximum temperature of the optically thin 
        thermal plasma.\\
        \hspace{1cm}
        $^{\ddagger}$Power of DEM as $d(EM) \propto (T/T_{\rm max})^{\rm \alpha-1} dT$.\\
        \hspace{1cm}
        $^{\S}$Solid angle of the reflector viewed from the plasma.\\
        \hspace{1cm}
        $^{\parallel}$Angle between the reflection surface normal and
 the observer's line of sight.\\
        \hspace{1cm}
        $^{\sharp}$Based on the solar abundances defined in \citet{1989GeCoA..53..197A}.\\
      }
\end{table}
All the errors are at 90\% confidence level. The reduced $\chi^2$ of
1.12 with 313 degree of freedom indicates that the fit is marginally
acceptable.
The maximum temperature and the power-law index of emission measure of
the plasma is obtained to be $T_{\rm max} = 33.5^{+7.1}_{-6.1}$ keV and
$\alpha = 0.58^{+0.19}_{-0.17}$, respectively. Our $T_{\rm max}$ is
consistent with that of \citet{2000MNRAS.315..307B},
$43_{-12}^{+13}$~keV using the Ginga data in which they adopted the same
model scheme to ours (see \S~\ref{sec:dis_mass} for more detail). The
best-fit $\alpha$ is consistent both with the value 0.5 predicted by
the isobaric post-shock plasma model
\citep{1994MNRAS.266..367I} and with 0.43
based on the full theoretical model
\citep{2005A&A...444..561F}.
The measured solid angle, 
$\Omega/2\pi=0.95^{+0.40}_{-0.27}$ can be interpreted as
unity, which suggests that the reflector is the WD surface and the shock
height is negligibly small compared with the WD radius.
It should be noted that the central energy of the iron K$\alpha$ line is
slightly smaller than its laboratory value 6.40~keV, and the profile is
significantly broad in energy.
The oxygen and iron abundances are measured at ${\rm Z_{\rm O}} =
1.07_{-0.39}^{+0.71}$ and ${\rm Z_{\rm Fe}} = 0.29\pm0.03$,
respectively.

Since our $\alpha$-free fit results in $\alpha =
0.58^{+0.19}_{-0.17}$ which is consistent with the theoretical value,
we fitted to the averaged spectra with the model fixing
$\alpha = 0.43$.  In this case, the reduced $\chi$-squared value is
$\chi^2_\nu = 1.12$ with 314 degree of freedom,
similar to that of the $\alpha$-free fit. 
$T_{\rm max}$ 
 calculated at larger value than that of the $\alpha$-free fit by 4.5 keV, 37.9$^{+5.1}_{-4.6}$ keV.
$\Omega/(2\pi)$ and $Z_{\rm O}$ are calculated at slightly
smaller values, $0.91\pm 0.26$ and
$0.86^{+0.40}_{-0.26}$, respectively, the other parameters almost common
with the $\alpha$-free fit. All parameters are
consistent with the $\alpha$-free case within their errors.

\subsection{Phase-resolved Spectra}\label{sec:phasespe}

We carried out phase-resolved spectral analysis by sorting the data
according to the WD spin phase.
First, we created spectra of the phases 0--0.1, 0.1--0.2, 0.2--0.3,
$\ldots$, 0.9--1.0 in Fig.~\ref{fig:flc} 
and calculated the ratios of these spectra to the phase-averaged
spectrum. The results are shown in Fig.~\ref{fig:ratio_spe}. 
\begin{figure}[h!]
  \begin{center}
      \FigureFile(80mm,50mm){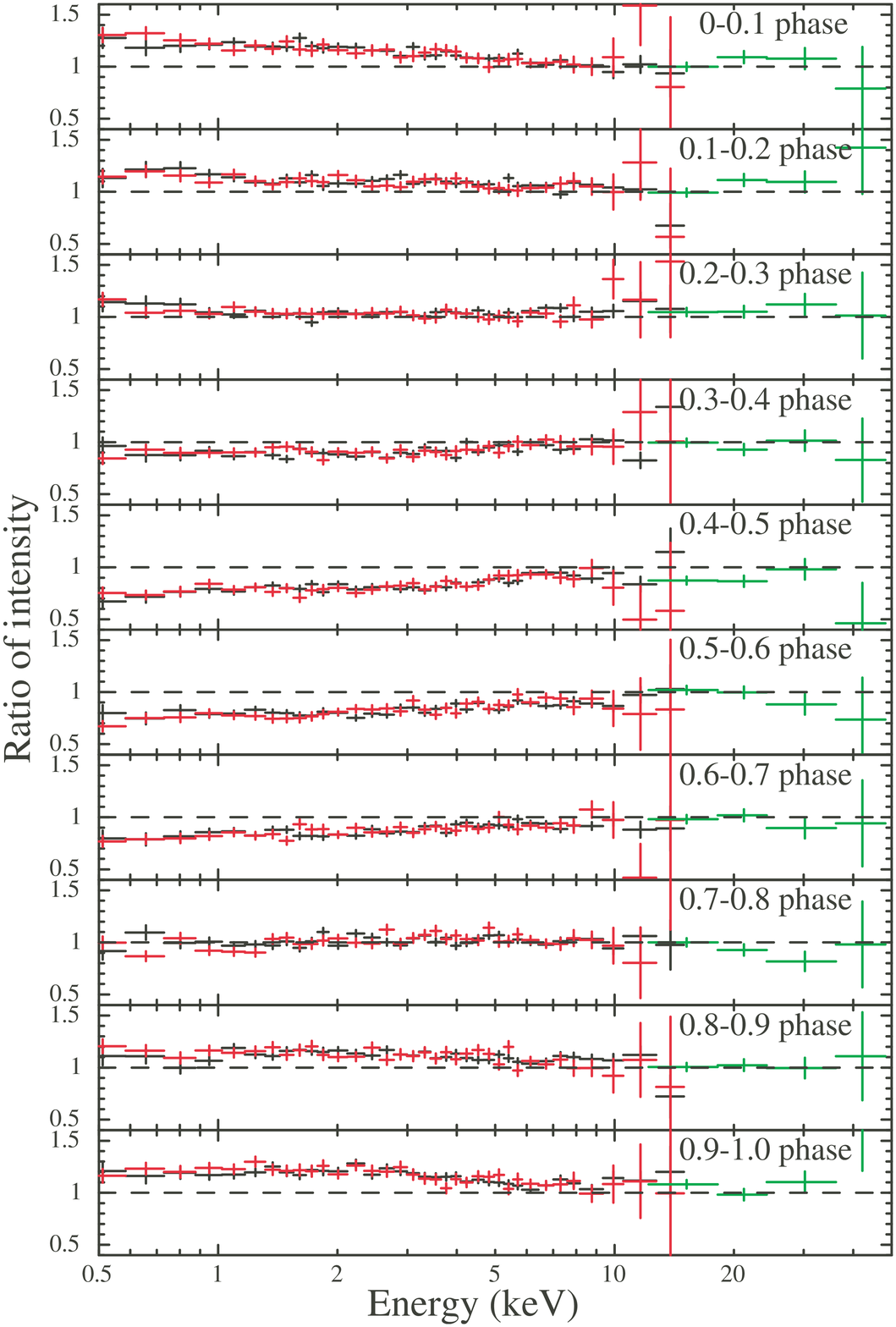}
  \end{center}
  \caption{Ratios of phase-resolved spectra to the averaged spectra. The
    spin cycle is segmented into 0.1 phase intervals, where the phase 0
    corresponds to the rotational intensity-maximum phase.  Black, red
    and green colors are the same as in Fig.~\ref{fig:avespe} and
    \ref{fig:avespe_fit}.} \label{fig:ratio_spe}
\end{figure}
Significant variation of the spectra is detected below 10~keV where the
photoelectric absorption can affects the spectra. By contrast, no
significant variation was detected above 10~keV. These results are as
expected from the energy dependence of the folded light curves in
Fig.~\ref{fig:flc}.

Second, we fitted the model defined in \S~\ref{sec:avespe} to the
spectra and evaluated physical parameters quantitatively in each phase.
Considering data statistics, we divided the spectra into
the following 8 phases:
0--0.25, 0.125--0.375, 0.25--0.5, 0.375--0.625, 0.5--0.75, 0.625--0.875,
0.75--1.0, 0.875--1.125 of Fig.~\ref{fig:flc}. Note that these phases
have some mutual overlap.
In the fitting, we fixed $\alpha$ at 0.43
\citep{2005A&A...444..561F}, which is obtained based on the full
theoretical calculation \citep{2005A&A...435..191S}. We also froze
$\Omega/2\pi$ equal to unity, which is consistent with the value from
the fit with $\alpha = 0.43$ (table~\ref{table:ave_para}).
We fixed $T_{\rm max}$, $Z_{\rm O}$, and $Z_{\rm Fe}$ at the values
obtained from the phase averaged spectral fitting where $\alpha$ was
fixed at 0.43 (table~\ref{table:ave_para}).
On the other hand, the inclination angle of the reflector $i$ 
was set free to vary because it should depend on the spin of the WD. The
5--50 keV band was used as the averaged spectral analysis.

In Fig.~\ref{fig:top_bottom_spe}, we showed the result of the fit
in the 0.875--1.125 phase as an example 
which covers the peak of the X-ray modulation. 
\begin{figure}
  \begin{center}
    \FigureFile(80mm,50mm){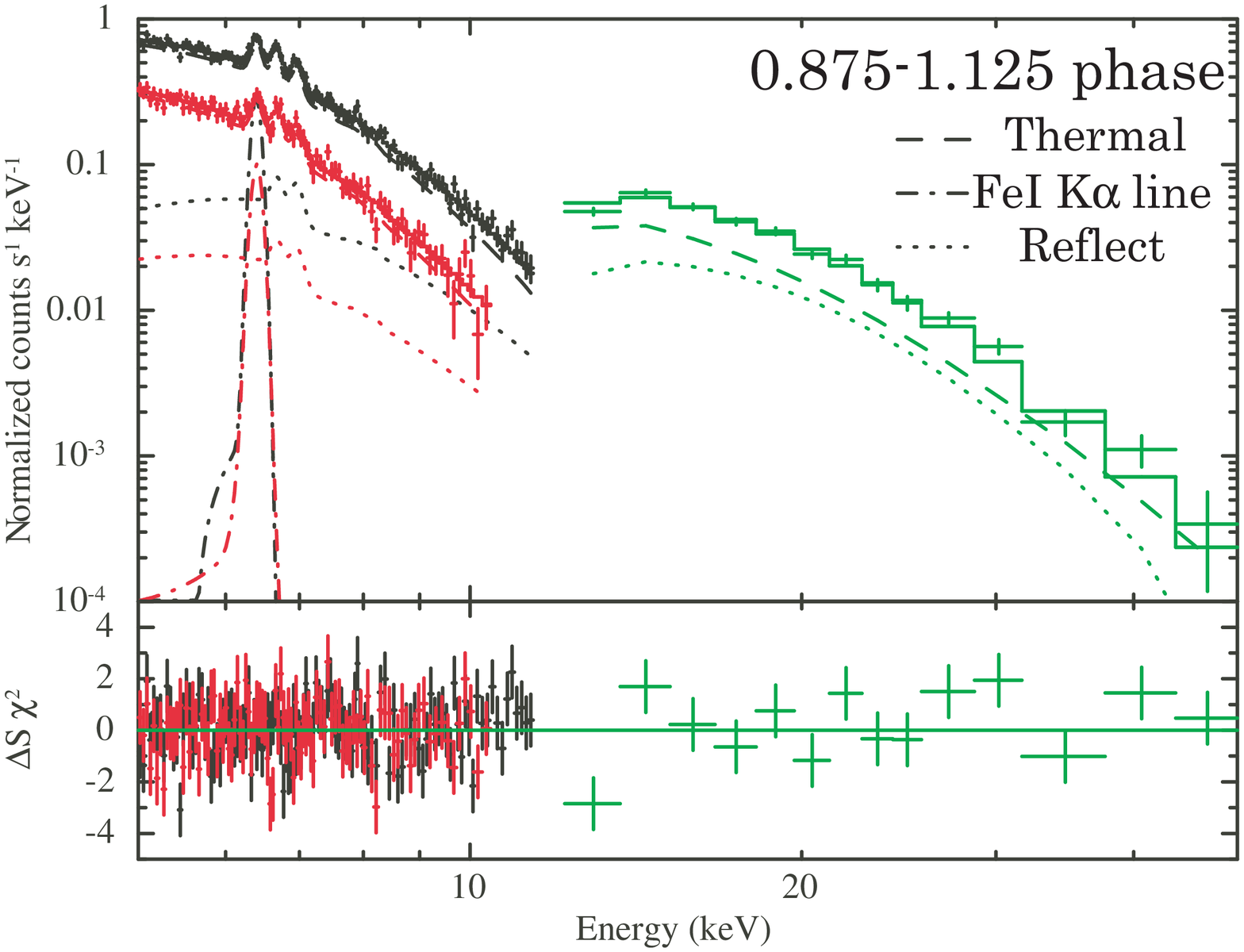}
  \end{center}
  \caption{Best-fit model to the spectra integrated over the 0.875-1.125
    phase of Fig.~\ref{fig:flc}. The color assignment is the same as in
    Fig.~\ref{fig:avespe} through \ref{fig:ratio_spe}.}
    \label{fig:top_bottom_spe}
\end{figure} 
The best-fit parameters are summarized in table
\ref{table:resolved_para} and are shown in Fig.~\ref{fig:phasecomp}.
\begin{figure}[h!]
  \begin{center}
    \FigureFile(80mm,50mm){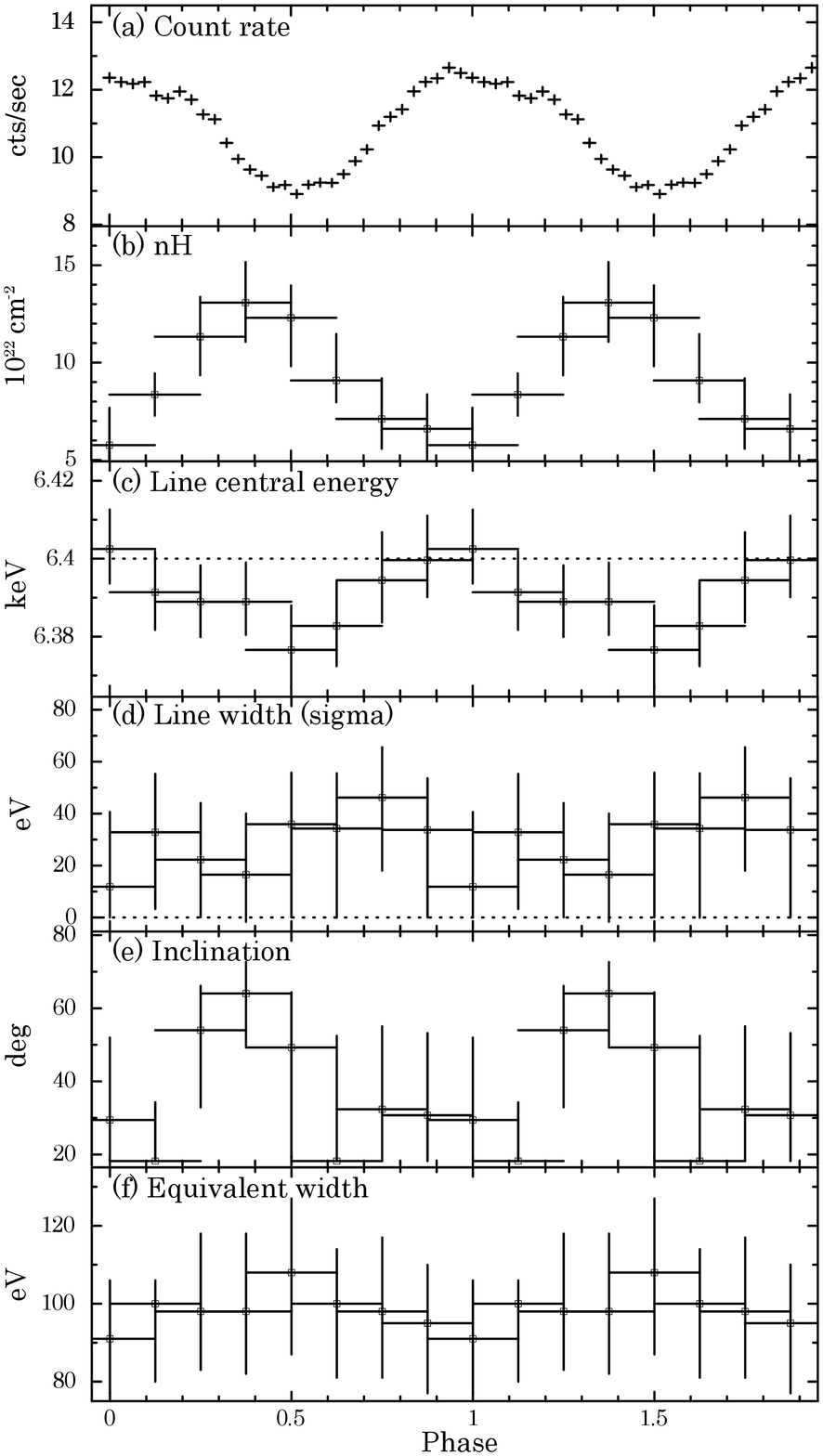}
  \end{center}
  \vspace{-0.5cm}
  \caption{The rotational modulation curves of (a) 0.1-11.5~keV X-ray count rate, 
  (b) the hydrogen column density N$_{\rm H}$, (c) the central energy of the fluorescent
 K$\alpha$ line, (d) the energy width of the Fe-I K$\alpha$ line, (e)
 the inclination of the reflector $i$, and (f) the equivalent width $EW$
 of the Fe-I K$\alpha$ line. Modulations of $N_{\rm H}$ and the center
 energy of the Fe-I K$\alpha$ line is significant.}
 \label{fig:phasecomp}
\end{figure} 
\begin{table}[h]
\rotcaption{Best-fit parameters of the phase resolved spectra.}\label{table:resolved_para} 
\centering
 \begin{sideways}
 \begin{tabular}{ccccccccc} 
 \\\hline
  Phase & 0.875-1.125 & 0-0.25 & 0.125-0.375 & 0.25-0.5 & 0.375-0.625 & 0.5-0.75 & 0.625-0.875 & 0.75-1\\\hline
  $N_{\rm H}$ ($\times10^{22}$ cm$^{-2}$) & $5.8_{-1.4}^{+1.9}$ & $8.4\pm1.1$ & $11.3\pm2.0$ & $13.1_{-2.0}^{+2.1}$ & $12.3_{-2.5}^{+1.7}$ & $9.1_{-1.1}^{+2.4}$ & $7.1_{-1.5}^{+2.1}$ & $6.6\pm1.8$\\
  T$_{\rm max}$$^{\ast}$ (keV) & 37.9 (fixed) & 37.9 (fixed) & 37.9 (fixed) & 37.9 (fixed) & 37.9 (fixed) & 37.9 (fixed) & 37.9 (fixed) & 37.9 (fixed)\\
  $\alpha^{\dagger}$ & 0.43 (fixed) & 0.43 (fixed) & 0.43 (fixed) & 0.43 (fixed) & 0.43 (fixed) & 0.43 (fixed) & 0.43 (fixed) & 0.43 (fixed)\\
  $\Omega/(2\pi)^{\ddagger}$ & 1 (fixed) & 1 (fixed) & 1 (fixed) & 1 (fixed) & 1 (fixed) & 1 (fixed) & 1 (fixed) & 1 (fixed)\\
 $i^{\parallel}$& $29_{-11}^{+23}$ & $\leq34$ & $54_{-21}^{+12}$ & $64_{-15}^{+ 9}$ & $49_{-30}^{+15}$ & $\leq52$ & $32_{-14}^{+23}$ & $31_{-13}^{+22}$\\
  ${\rm Z_{\rm O}}^{\S}$ & 0.86 (fixed) & 0.86 (fixed) & 0.86 (fixed) & 0.86 (fixed) & 0.86 (fixed) & 0.86 (fixed) & 0.86 (fixed) & 0.86 (fixed)\\
  ${\rm Z_{\rm Fe}}^{\S}$& 0.28 (fixed)  & 0.28 (fixed) & 0.28 (fixed) & 0.28 (fixed) & 0.28 (fixed) & 0.28 (fixed) & 0.28 (fixed) & 0.28 (fixed)\\
  FeI-K$\alpha$ line Energy (keV) & $6.403_{-0.009}^{+0.010}$ & $6.391_{-0.010}^{+0.011}$ & $6.389\pm0.009$ & $6.389_{-0.009}^{+0.010}$ & $6.377_{-0.010}^{+0.011}$  & $6.383_{-0.010}^{+0.011}$  & $6.394_{-0.011}^{+0.012}$  & $6.400_{-0.009}^{+0.011}$\\
 FeI-K$\alpha$ line sigma (eV) & $12_{-12}^{+29}$  & $33_{-30}^{+22}$  & $22_{-22}^{+22}$  & $16_{-16}^{+24}$  & $36_{-36}^{+20}$  & $34_{-34}^{+21}$  & $46_{-28}^{+19}$  & $34_{-34}^{+20}$\\
 FeI-K$\alpha$ line $EW$ (eV) & $91_{-14}^{+15}$  & $100_{-20}^{+6}$  & $98_{-15}^{+20}$  & $98_{-16}^{+20}$  & $108_{-21}^{+19}$  & $100_{-19}^{+14}$  & $98_{-17}^{+19}$  & $95_{-18}^{+15}$\\
 $\chi^2$ (d.o.f.) & 262.58 (226)  & 248.00 (226)  & 244.01 (226)  & 241.27 (226)  & 217.05 (226)  & 265.77 (226)  & 205.86 (226) & 246.63 (226)\\\hline
  \multicolumn{8}{l}{\footnotesize 
    $^{\ast}$Maximum temperature of the post-shock plasma.}\\
  \multicolumn{8}{l}{\footnotesize 
    $^{\dagger}$Power index of DEM as $d{EM} \propto (T/T_{\rm
  max})^{\alpha -1}dT$.}\\
  \multicolumn{8}{l}{\footnotesize 
    $^{\ddagger}$Solid angle of the reflector subtending over the
  post-shock plasma.}\\
  \multicolumn{8}{l}{\footnotesize 
    $^{\S}$Based on the solar abundances defined in \citet{1989GeCoA..53..197A}.}\end{tabular}
\end{sideways}
\end{table}
The fits are generally acceptable at the 90\% confidence level. The
parameter that shows the most significant spin modulation is hydrogen
column density $N_{\rm H}$, as demonstrated in panel (b),
which varies by up to a factor of two among the phases and is inversely
correlated with the X-ray light curve. The modulation
of $i$ is not clear. Although $i$ seems larger at around the
intensity-minimum phase, the statistical errors are somewhat too large
as shown in panel (e).
The central energy of the fluorescent iron K$\alpha$ line shown in panel
(c) shifts significantly from 6.40~keV to maximally
6.38~keV at the intensity-minimun phase, while its width does not change
within statistics errors (panel (d)). The $EW$ (panel (f)) may be
anti-correlated with the X-ray modulation, though it is critical
statistically.

At the end of this subsection, we would like to remark
on the central energy shift of the 6.4~keV line at the intensity-minimum
phase in Fig.~\ref{table:resolved_para}(c). It is wellknown that the
line central energy is sensitive to the evaluation of the slope of
underlying continuum. Accordingly, we have simulated the phase-resolved
spectra with the {\tt fakeit} command in XSPEC by varying $N_{\rm H}$ in
the range 1--50$\times 10^{22}$~cm$^{-2}$. In doing this, we used XIS0,
1, 3 and PIN and created combined spectra in the 5-50~keV band with
exposure times equal to those of the phase-resolved spectra. The
parameters except $N_{\rm H}$ are fixed at the best-fit values
(table\ref{table:resolved_para}). As a result of the fits, we found that
the shift of the line central energy is 5~eV at the phase 0.375-0.625
where the central energy shift of 6.4 keV line is maximum.
We therefore conclude that the line central energy shift of $\sim$20~eV at
the rotational intensity minimum phase is real.

\section{Discussion}\label{sec:dis}

\subsection{Estimation of mass of WD and comparison with the other observations}\label{sec:dis_mass}
As explained in \S~\ref{sec:intro}, the temperature at the shock front $T_{\rm max}$ 
is represented with Eq.~(\ref{eq:shockT}).
The plasma encountered the shock cools via optically thin
thermal plasma emission and finally settles onto the WD.  Note that the
cyclotron radiation important in polars, which harbor a WD with a
magnetic field of as strong as $B \simeq 10^{7-9}$~G
\citep{1996A&A...306..232W},
can be ignored in {\vIZ}. With the aid of the WD mass and radius
relation by \citet{1972ApJ...175..417N},
\begin{eqnarray}
  \hspace{-0.5cm}
  R_{\rm WD} = 0.78 \times 10^9
  \left[ \left(\frac{1.44\MO}{M_{\rm WD}}\right)^{2/3}
    -\left(\frac{M_{\rm WD}}{1.44\MO}\right)^{2/3}\right] ^{1/2}{\rm cm}
  \hspace{-1cm}
  \label{eq:M_R}
\end{eqnarray}
and the observed $T_{\rm max}$ calculated with fully theoretical $\alpha$ = 0.43
in Eq.~(\ref{eq:shockT}), we obtained the
WD mass and the radius to be 
$M_{\rm WD} = 0.82^{+0.05}_{-0.06}~\MO$ and $R_{\rm WD} = (6.9\pm0.4)\times10^{8}$~cm
, respectively.

\begin{longtable}{*{4}{c}}
  \caption{Collection of WD mass estimation in the {\vIZ}.}\label{table:mass}
  \hline
  Telescope  & Energy band (keV)& $M_{\rm WD} (\MO)$ & Reference\\\hline
  \endfirsthead
  \hline
 \endhead
 \hline
 \endfoot
 \hline
 \endlastfoot
  Ginga & 2-20 & $0.93 \pm 0.12$ & \citet{2004A&A...419..291B}\\
  RXTE $\&$ INTEGRAL & 3-100 & $0.71 \pm 0.03$ & \citet{2004A&A...426..253R}\\
  RXTE & 3-80 & $0.95 \pm 0.05$ & \citet{2005A&A...435..191S}\\
  XMM-Newton & 0.2-10 & $1.046^{+0.049}_{-0.012}$ & \citet{2007ApJ...663.1277E}\\
  Swift & 15-100 & $0.65 \pm 0.04$ & \citet{2009A&A...496..121B}\\
  Suzaku & 4-50 & $0.75\pm0.05$ & \citet{2010A&A...520A..25Y}\\
  Suzaku & 5-50& $0.82^{+0.05}_{-0.06}$ & This work
 \end{longtable}

A number of publications have been made so far on the mass determination of
the WD in {\vIZ}. The results of them are summarized in
table~\ref{table:mass}.
As evident from this table, they show a large scatter, amounting to $\sim$0.4\MO.
Of them, \citet{2005A&A...435..191S}, \citet{2007ApJ...663.1277E},
\citet{2009A&A...496..121B} used data of RXTE, XMM-Newton, and Swift,
respectively, and evaluated the mass by taking into account the
multi-temperature nature of the post-shock plasma.
They, however, do not include the reflection component in their spectral
evaluation.
Its intensity relative to the direct component, however, increases with
increasing X-ray energy due to predominance of the Compton scattering (=
Thomson scattering) over the photoelectric absorption. In still higher
energies, on the contrary, the reflection component becomes relatively
weaker because of reduction of the Compton cross section with increasing
X-ray energy. The boundary between these two phenomena appears at
$\sim$25~keV, as demonstrated in Fig.~\ref{fig:avespe_fit} and
\ref{fig:top_bottom_spe}. Since the energy band of RXTE and XMM-Newton
is effectively limited below $\sim$30~keV, the omission of the
reflection component leads \citet{2005A&A...435..191S} and
\citet{2007ApJ...663.1277E} to a $T_{\rm max}$ evaluation larger than
reality. In fact, we fitted our spectra without the reflection component
in the 5--30~keV band, and found $kT_{\rm max} > 100$~keV.
\citet{2009A&A...496..121B}, on the other hand, evaluated $kT_{\rm max}$
with the Swift data in the 15--100~keV energy band where the reflection
component softens spectra. Accordingly, their $kT_{\rm max}$ is lower
than as it is, resulting in a smaller estimation of the WD mass. Since
the statistics of our HXD-PIN spectrum is only moderate, we have
simulated the HXD-PIN spectrum with the {\it fakeit} command in XSPEC
using the best-fit parameters (\S~\ref{sec:avespe}), and evaluated it
with the model but without the reflection component in the band
15--60~keV. The maximum temperature results in as low as 16~keV.

{\citet{2004A&A...426..253R}} took account of the reflection component,
as well as the multi-temperature nature of the post-shock plasma. Their
estimation of the reflector's solid angle, $\Omega/2\pi = 0.35\pm0.15$,
was smaller than ours by a factor of a few. 
\citet{2004A&A...419..291B} used the plasma temperature of $kT_{\rm max}
= 43^{+13}_{-12}$ keV estimated by \citet{2000MNRAS.315..307B} from
Ginga. \citet{2000MNRAS.315..307B} took the multi-temperature nature
of the plasma into consideration and the reflection which assumed that
the reflector occupy the solid angle of $\Omega/2\pi = 1$, which
coincides with our result. Accordingly, their mass range overlaps with
ours within statistical error.

\citet{2010A&A...520A..25Y} adopted a partial covering absorption model,
characterized with a covering fraction of $C_{\rm PC} = 0.5^{+0.04}_{-0.02}$ and 
a very large hydrogen column density of $n_{\rm H} = 219^{+29.7}_{-25.7}\times10^{22}$cm$^{-2}$,
instead of the reflection component.
\citet{1998MNRAS.293..222C} revealed that a partial covering model
can mimic the reflection model reasonably well, and the WD mass only slightly change from
an estimation based on the reflection model.
In fact, their estimation of WD mass of $0.75\pm0.05~\Mo$ matches with ours.

It is therefore important to remember that the contribution of the
reflection component must be taking into account correctly for evaluating the mass
of the WD. Usage of the HXD-PIN onboard Suzaku is one of the best ways
to do so in that it has the highest sensitivity at the turnover
energy $\sim$25~keV among all the instruments so far in orbit.

\subsection{The Nature of the Fluorescent Iron K$\alpha$ Line}
\label{sec:dis_am}

\subsubsection{Origin of the Energy Shift of the Line}\label{sec:origin_shifted_line}

As demonstrated in Fig.~\ref{fig:phasecomp} and
table~\ref{table:resolved_para}, we detected spin-modulation of the
central energy of the fluorescent iron K$\alpha$ line that becomes
minimum at the spin-minimum phase. This characteristic can be understood in
the framework of the
accretion curtain model \citep{1988MNRAS.231..549R} accepted widely
today, in which the X-ray intensity modulation is caused mainly by
photoelectric absorption in the pre-shock accretion column. The spin-minimum
phase occurs when the upper accretion column points to the observer, at
which configuration the observer looks down the post-shock plasma
through the pre-shock absorbing matter. At the peak phase, on the other
hand, the absorption effect is minimum, and the second pole may appear
depending on the inclination angle. The observed line-of-sight hydrogen
column density (panel (a) and (b) of Fig.~\ref{fig:phasecomp}) behaves according to this
picture. The observed red-shifted fluorescent iron line component can be
attributed to the pre-shock accreting matter which flows away from the
observer to the WD. \citet{1999ApJS..120..277E} pointed out based on
ASCA observations of $\sim$20 mCVs that the fluorescent iron line mainly
originates from the WD surface, and indicated significant contribution
from the pre-shock accretion column as well if the thickness of the
accretion column is $\gtsim$10$^{23}$~cm$^{-2}$. Since the rotational
speed of the WD surface (= $2\pi R_{\rm WD}/P_{\rm spin}$) is  
60~km~s$^{-1}$, the energy shift of the line associated with the WD rotation
amounts only to $\sim$1~eV. Consequently, the energy shift of the iron line is most
likely attributed to the contribution from the pre-shock accretion column.

\subsubsection{Identification of the Fluorescent Iron K$\alpha$ Line
   from the Pre-shock Accretion Column}
To further confirm the fluorescence Fe line emission originating from the pre-shock
accretion column, we looked into the phase-resolved spectra in the
5--8.5~keV band in more detail. First, we fitted the same model as in
\S~\ref{sec:phasespe} but with the iron line central energy and its width
being fixed at 6.4~keV and 0~keV, respectively. We fixed all the
continuum parameters at their best-fit values (table \ref{table:resolved_para}), but only the
normalization is set free to vary.
The top and middle panels of Fig.~\ref{fig:1ga_2ga} are the results of
the fits to the spectra at the spin-maximum and the spin-minimum phases 
(0.875-1.125 and 0.375-0.625, respectively, see Fig.~\ref{fig:flc}) of the X-ray
intensity modulation.
\begin{figure}[h!]
  \begin{center}
    \begin{minipage}{0.95\hsize}
      \begin{center}
        \FigureFile(80mm,50mm){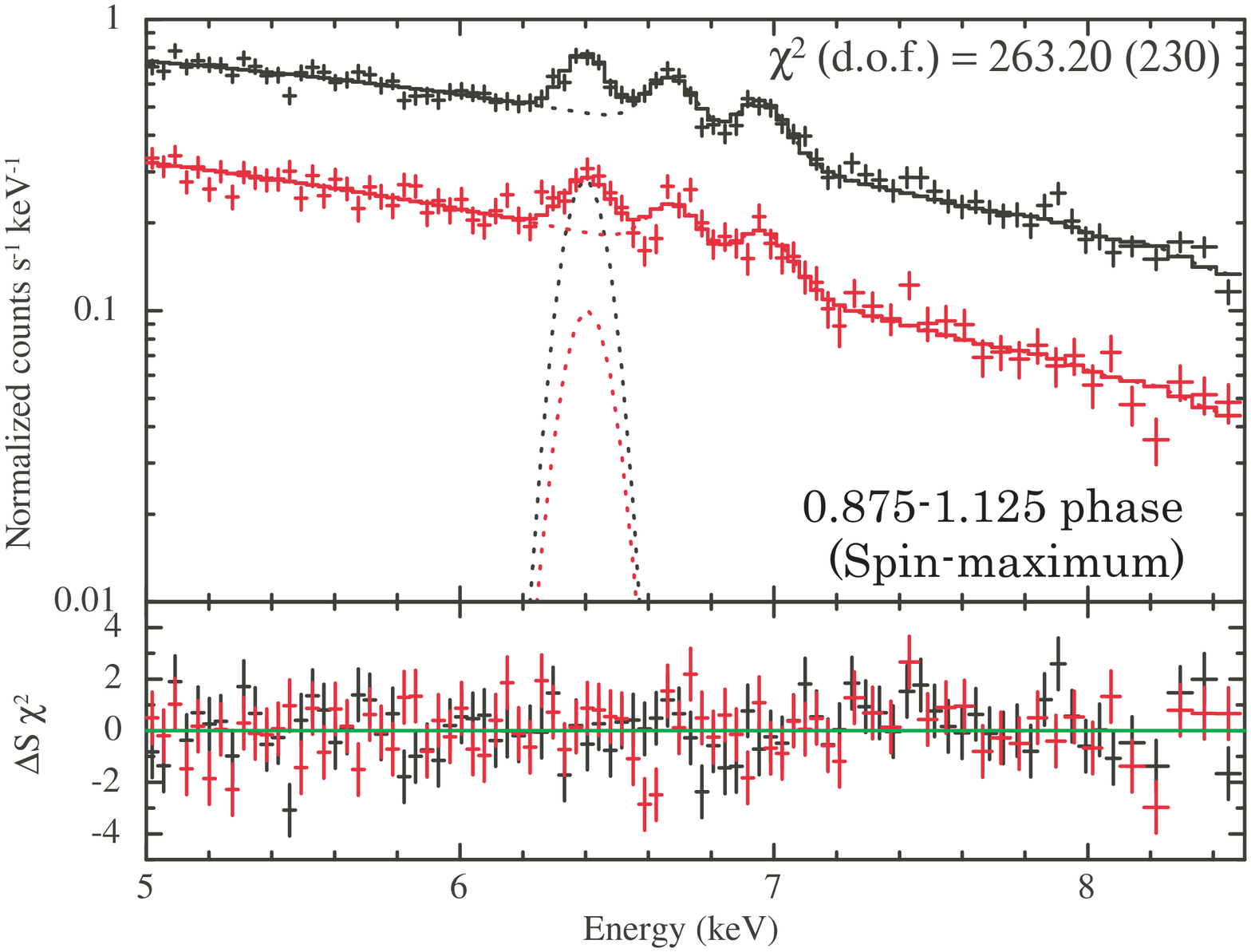}
      \end{center}
    \end{minipage}
    \begin{minipage}{0.95\hsize}
      \begin{center}
        \FigureFile(80mm,50mm){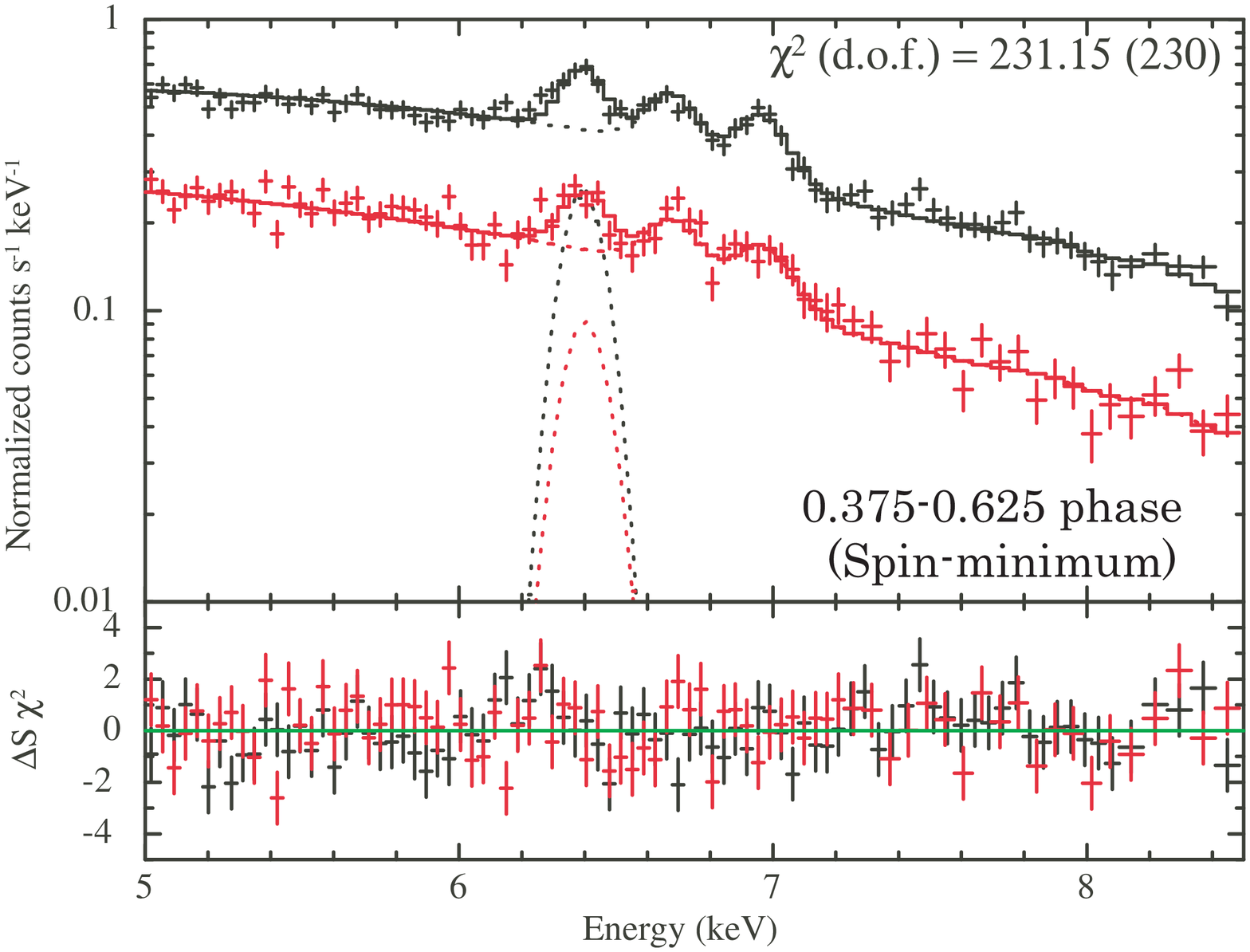}
      \end{center}
    \end{minipage}
    \begin{minipage}{0.95\hsize}
      \begin{center}
        \FigureFile(80mm,50mm){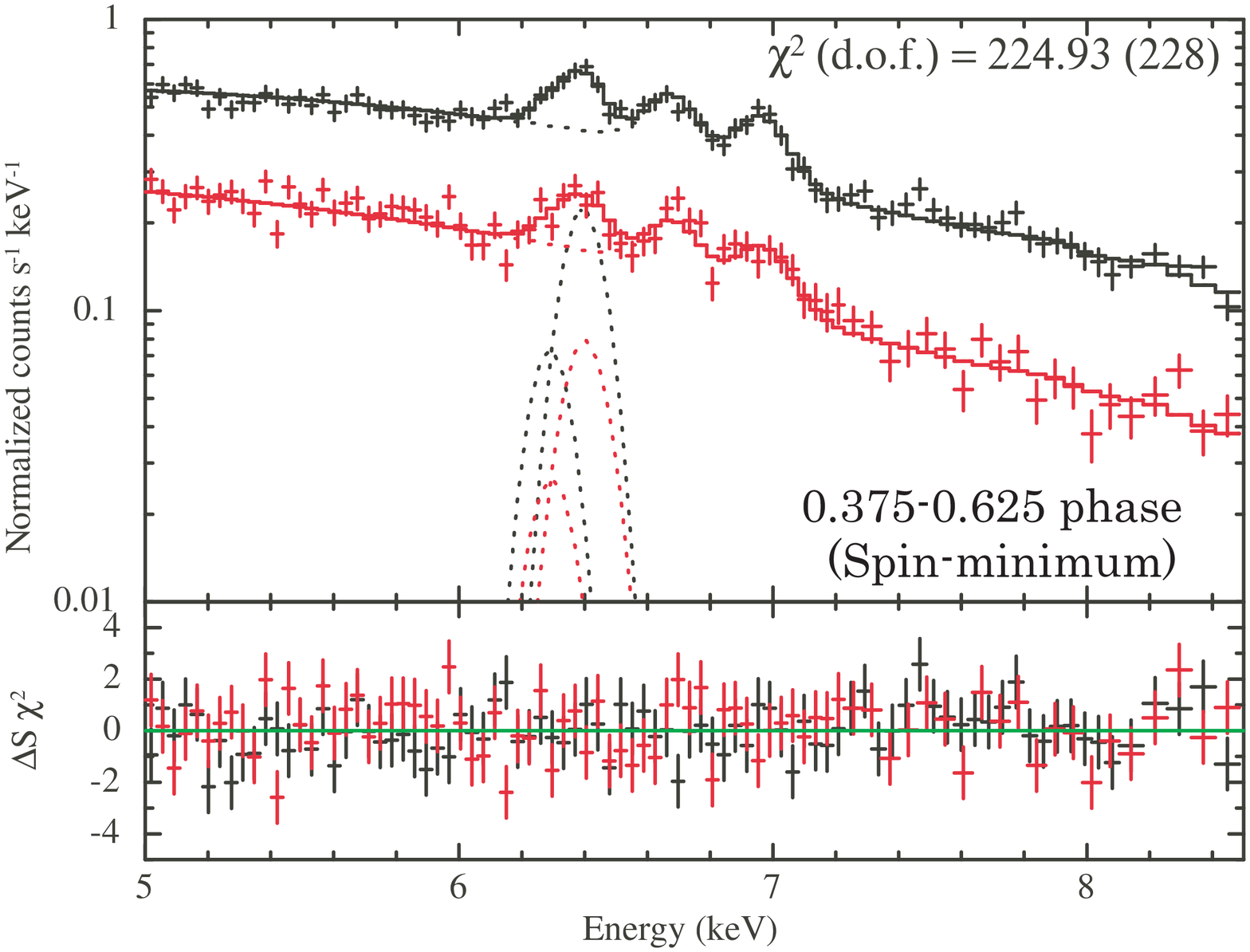}
      \end{center}
    \end{minipage}
  \end{center}
  \vspace{-0.5cm}
  \caption{The 5-8.5 keV spectra fitted with a single Gaussian at the
 intensity-peak phase 0.875-1.125 (top) and the intensity-minimum phase
 0.357-0.625 (middle). The with of the Gaussian is fixed at $\sigma$ =
 0~keV. (Bottom) The same as the middle panel but fitted with two
 Gaussians.} \label{fig:1ga_2ga}
\end{figure} 
In the  spin-maximum phase, it is clear that the structure of the line can be
reproduced well only with a single narrow Gaussian. In the spin-minimum phase,
on the other hand, a slight wiggle remains centered at $\sim$6.25~keV in
the residual panel, although the fit itself is statistically acceptable
($\chi^2$ (d.o.f.) = 231.15 (230)).

In order to fill the residual, we added another Gaussian in the model
and fitted it to the phase-resolved spectra. The best-fit parameters are
summarized in table~\ref{table:para_2ga}.
\begin{table}[h]
\rotcaption{Best-fit parameters of the phase resolved spectra with two Gaussians.}\label{table:para_2ga}
\centering
 \begin{sideways}
 \begin{tabular}{ccccccccc} 
 \\\hline
  Phase & 0.875-1.125 & 0-0.25 & 0.125-0.375 & 0.25-0.5 & 0.375-0.625 & 0.5-0.75 & 0.625-0.875 & 0.75-1\\\hline
  FeI-K$\alpha$ line Energy$^{\dagger}$ (keV) & 6.4 (fixed) & 6.4 (fixed) & 6.4 (fixed) & 6.4 (fixed) & 6.4 (fixed) & 6.4 (fixed) & 6.4 (fixed) & 6.4 (fixed)\\
  FeI-K$\alpha$ line sigma$^{\dagger}$ (eV) & 0 (fixed) & 0 (fixed) & 0 (fixed) & 0 (fixed) & 0 (fixed) & 0 (fixed) & 0 (fixed) & 0 (fixed)\\
  FeI-K$\alpha$ line $EW$$^{\dagger}$ (eV) & $90_{-9}^{+8}$ & $95_{-7}^{+11}$ & $86_{-80}^{+19}$ & $91_{-91}^{+14}$ & $80_{-51}^{+13}$ & $83_{-44}^{+13}$ & $92_{-7}^{+11}$ & $92_{-7}^{+11}$\\
    FeI-K$\alpha$ line Energy$^{\ddagger}$ (keV) & - & - & $6.33_{-0.22}^{+0.07}$ & $6.29_{-0.11}^{+0.10}$ & $6.30_{-0.05}^{+0.07}$ & $6.31_{-0.06}^{+0.05}$ & - & -\\
  FeI-K$\alpha$ line sigma$^{\ddagger}$ (eV) & 0 (fixed) & 0 (fixed) & 0 (fixed) & 0 (fixed) & 0 (fixed) & 0 (fixed) & 0 (fixed) & 0 (fixed)\\
  FeI-K$\alpha$ line $EW^{\ddagger}$ (eV) & - & - & $25_{-23}^{+90}$ & $13_{-11}^{+102}$ & $28_{-13}^{+44}$ & $24_{-15}^{+71}$ & - & -\\
$\chi^2$ (d.o.f.) & 263.20 (230) & 210.09 (230) & 243.91 (228) & 240.27 (228) & 224.93 (228) & 264.75 (228) & 245.60 (230) & 247.92 (230)\\
  Significance $^{\S} (\%)$ & - & - & 83.5 & 83.9 & 95.5 & 97.5 & - & - \\\hline
  \multicolumn{5}{l}{\footnotesize 
    $^{\dagger}$The component stable in energy.}\\
  \multicolumn{5}{l}{\footnotesize 
    $^{\ddagger}$The red-shifted component.}\\
  \multicolumn{5}{l}{\footnotesize 
    $^{\S}$Significance of $F$-test for adding a second Gaussian to
  fluorescent iron K$\alpha$ line}\\
\end{tabular}
\end{sideways}
\end{table}
Here the width of the second Gaussian is fixed at 0~keV. As suggested
from Fig.~\ref{fig:1ga_2ga}, the second Gaussian is found not required
at some phases statistically, and hence, we added it only to the phases
where significance of the $F$-test exceeds 80, whose values are listed
at the bottom line of table~\ref{table:para_2ga}. As evident from the
table, introduction of the second Gaussian significantly improves the
fits at phases close to the spin-minimum phases.
 The shift of the central energy of the second Gaussian is the largest at the phase 0.25-0.5 where
the central energy becomes 6.29~keV. The $EW$ of the component is
approximately 30~eV while the uncertainties are somewhat large. 
The bottom panel of Fig.~\ref{fig:1ga_2ga} shows the XIS spectra fitted
with the model including the two Gaussians, as an example. The wiggle in
the residual plot of the top panel disappears. 

\subsubsection{Origin of the Red-shifted Iron K$\alpha$ Line}

At the spin-minimum phase, we have succeeded in resolving the red-shifted
fluorescent iron K$\alpha$ line whose central energy is
$6.30^{+0.07}_{-0.05}$~keV
at the phase 0.375-0.625. The corresponding
line-of-sight velocity is
$4.7^{+2.3}_{-3.3}\times10^3$
~km~s$^{-1}$. This agrees well with the
free-fall velocity onto the WD $= \sqrt{\mathstrut 2GM_{\rm WD}/R_{\rm WD}}$
= $5.6\pm0.4$ $\times10^3$~km~s$^{-1}$. Since the pre-shock matter is expected
to flow toward the WD at the free-fall speed, and to be maximally
red-shifted at the spin-minimum phase, the red-shifted iron line must
originate from the pre-shock accreting matter via fluorescence due to
irradiation from the post-shock plasma. The WD surface velocity can not explain the
observed velocity shift (\S~\ref{sec:origin_shifted_line}).
In the WD rotation picuture, relative phasing of the line energy shift
with respect to the X-ray intensity modulation is somewhat different
from what is observed, and at some phases a blue-shifted component must
have been detected.

\subsubsection{Verification with equivalent width}

The $EW$ of the red-shifted iron line has been estimated to be $EW$ =
$28^{+44}_{-13}$ eV at the 0.375-0.625 phase.
The $EW$ of fluorescent iron K$\alpha$ line
from point-like source surrounded by matter with a column density of
$N_{\rm H}$ by a 2$\pi$ solid angle
\begin{eqnarray}
\label{eq:EW_inoue}
EW =\frac{1}{2}\,\beta\,\frac{N_{\rm H}}{10^{21}\hspace{0.5 ex}
{\rm cm^{-2}}}\hspace{0.5 ex}
\hspace{0.5 ex}\;\;\;\;\mbox{[eV]}
\end{eqnarray}
\citep{1999ApJS..120..277E}. The factor 1/2 is introduced because we
consider fluorescence of the accreting matter and hence the hemisphere
not occupied by the WD. $\beta$ is the correction factor of spectral
shape of an ionizing continuum, and $\beta = 1$ if the continuum is a
power law with a photon index of 1.1 \citep{1985SSRv...40..317I}.
The $EW$ scales with power of continuum to ionize
iron K-shell, and can be described with the ionizing spectrum
$f_{\rm I}$ as
\begin{eqnarray}\label{eq:correction}
EW \propto \frac{1}{f_{\rm I}(E_{\rm K\alpha})}
\int^{\infty}_{E_{\rm K\hspace{0.5 ex}edge}}
f_{\rm I}(E)\sigma_{\rm Fe}(E_{\rm K\hspace{0.5 ex}edge}, E)dE,
\label{eq:correct}
\end{eqnarray}
where E$_{\rm K~edge}$ and E$_{\rm K\alpha}$ are the energies of the
iron K edge and the central energy of the neutral iron K$\alpha$ line,
which are 7.1~keV and 6.4~keV, respectively. $\sigma_{\rm Fe}(E_{\rm
K~edge}, E)$ is the photo ionization cross section of iron K-shell
\citep{1990A&A...237..267B}. $\beta$ can be obtained by the ratio of
Eq.~(\ref{eq:correct}) calculated with the multi-temperature plasma
spectrum to that calculated with a power law with $\Gamma = 1.1$.
In our case, the multi-temperature plasma with $kT_{\rm max} =$ 37.9~keV
results in $\beta = 0.73$. 
The hydrogen column density at the spin-minimum phase 
has been evaluated $N_{\rm H}$ = $12.3^{+1.7}_{-2.5}\times10^{22}$
cm$^{-2}$ (table~\ref{table:resolved_para}).
Note that correction of the iron abundance is not necessary in Eq.~(\ref{eq:EW_inoue})
because both the equivalent width and $N_{\rm H}$ scale in proportion to the iron abundance.
Consequently, the $EW$
of the fluorescent iron K$\alpha$ line from the pre-shock accreting
matter is expected to be $EW$ = 44.9$^{+6.2}_{-9.1}$ eV.
This value agrees with the observed one 
$28^{+44}_{-13}$ eV.

We note that we have to adopt $N_{\rm H}$ in Eq.(\ref{eq:EW_inoue})
which is averaged over the hemisphere not occupied by the white
dwarf. It is, however, impossible to evaluate the averaged $N_{\rm H}$
because we can measure the thickness of matter on the line of sight. As
an approximation, we might have used the phase-average value listed in
table~\ref{table:para_2ga}, since we can observe the post-shock plasma
from different directions using the WD rotation. However, line-of-sight
$N_{\rm H}$ varies in a small range
0.58-1.31$\times 10^{23}$~cm$^{-2}$. Accordingly, it does not matter to use the $N_{\rm
H}$ at the spin-minimum phase as a representative, considering the large error
on the observed $EW$.

\bigskip

In summary of this subsection, we detected the iron K$\alpha$ line
component whose central energy is modulated synchronized with the WD
rotation. Its modulation pattern (maximal energy shift at the X-ray
intensity minimum), amount of energy shift, and $EW$ all indicate that
the line originates from the pre-shock accreting flow via fluorescence
due to irradiation of the post-shock hot plasma emission.

\subsection{Height and fractional area of the post-shock accretion
  column}
\label{sec:Geo_AC}

The intensity of the reflection continuum and the fluorescent iron
K$\alpha$ emission line at 6.4~keV from the WD surface are good indicator
of the height of the post-shock plasma since they both depend upon the
solid angle of the plasma viewed from the WD. In this subsection we
intend to constrain the plasma height from these observed quantities,
and then try to evaluate the size of the fractional accreted area on the
WD surface.

As mentioned in \S~\ref{sec:avespe}, the solid angle of the WD surface is
measured to be $\Omega/2\pi=0.91\pm0.26$ through the spectral
fitting to the averaged spectra. The result is consistent with $\Omega =
2\pi$, implying that the height of the post-shock plasma is negligibly
small compared with the WD radius. The upper limit of the height
$h_\Omega$ is obtained from the lower limit of $\Omega$, which results
in $h_\Omega < 0.07~R_{\rm WD}$, assuming that the post shock plasma is
point-like.

A similar estimation can be carried out using the $EW$ of the 6.4~keV
iron line in the framework of \citet{1991MNRAS.249..352G}, in which they
evaluated the 6.4~keV line $EW$ in a geometry composed of a point-like
X-ray source locating above an infinite optically thick slab.
In their scheme, \citet{1991MNRAS.249..352G} assumed that the
irradiating spectrum from the point-like source is a power-law with
various photon indices. They assumed the iron abundance of the slab to be [Fe/H] =
$3.2\times10^{-5}$. The X-ray spectrum of {\vIZ} is, on the other hand,
a multi-temperature plasma emission with the maximum temperature of
37.9~keV. Since the temperature is very high, the spectrum can be
approximated with a power law at energies which ionize a K-shell
electron of iron efficiently ($E \gtsim E_{\rm K~edge} = 7.1$~keV). 
To identify the power-law model that represents the observed continuum best, 
we produced a simulated XIS spectrum based on the
multi-temperature plasma emission model which best-fits the
phase-averaged spectra (Fig.~\ref{fig:avespe_fit} and
table~\ref{table:ave_para}) using the {\tt fakeit} command in XSPEC, and
fitted it with a power law.
As a result, we found the irradiating spectrum can be approximated with
a power law with $\Gamma$ = 1.46.
Our estimation of the iron abundance 0.28
is based on the assumption of [Fe/H] = $4.68 \times 10^{-5}$. In the
scheme of \citet{1991MNRAS.249..352G}, we should use the value $Z_{\rm
Fe} = 0.41$ after correction of the abundance difference in solar
abundance.

After all these manipulations, the $EW$ expected from {\vIZ} in the case
of negligible height plasma is obtained to be $\sim$80~eV, which is
consistent with the observed $EW = 80^{+13}_{-51}$~eV at the spin-minimum
 0.375--0.625 phase (table~\ref{table:para_2ga}).
Consequently, the $EW$ of fluorescent iron K$\alpha$ line is also
consistent with a picture in which the height of the plasma is
negligibly small.
The upper limit of the plasma height $h_{\rm K\alpha}$ estimated from
the lower bound of the observed $EW$ 
(= 29~eV) is $h_{\rm K\alpha} < 0.15~R_{\rm WD}$. 
Adopting the limit on $h_{\rm \Omega}$,
we conclude that the height of
the post-shock plasma is less than 7~\% of the WD radius.

In a one-dimensional accretion flow model dominated by bremsstrahlung
cooling \citet{1973PThPh..50..344A}, which can be applied to {\vIZ},
the height of the plasma can be analytically solved as
\begin{equation}
h_c = 0.605v_st_c,
\label{eq:x}
\end{equation}
where $v_s$ is the velocity of the post-shock plasma at the shock front,
which is 
\begin{eqnarray}
v_s & = & \frac{1}{4} \sqrt{\mathstrut {\frac{2GM_{\rm WD}}{R_{\rm
 WD}}}} \nonumber \\
& = & 1.40\times 10^8\;\;\;\mbox{[cm s$^{-1}$]}
\label{eq:vb}
\end{eqnarray}
in the case of a strong shock, where we adopt $M_{\rm WD} = 0.82~\Mo$ and
$R_{\rm WD}= 6.9\times 10^8$~cm, obtained in this work
(\S~\ref{sec:dis_mass}). $t_c$ is the cooling time scale at the shock
front
\begin{eqnarray}
t_c & = & 3kT_{\rm max}/(2\mu m_H\epsilon_{f\!\!f}) \nonumber \\
 & = & 4.20\times10^{-13}T^{1/2}\rho_s^{-1}\;\;\;\mbox{[s]},
\label{eq:tc}
\end{eqnarray}
in the c.g.s. unit, where $\varepsilon_{f\!\!f}$ is the bremsstrahlung
emissivity with the Gaunt factor of 1.10 from Born approximation, and
$\rho_s$ is the density of the post-shock plasma at the shock front,  
which can be described as
\begin{equation}
\rho_s = \frac{\dot{M}}{4\pi fR_{\rm WD}^2 v_s},
\label{eq:rho}
\end{equation}
Of the quantities in Eq.~(\ref{eq:rho}), $\dot{M}$ can be evaluated from
the observed bolometric flux $F_{0.1-100} = 4.0\times10^{-10}$ erg
cm$^{-2}$ s$^{-1}$ together with $D = 527$~pc
\citep{2004A&A...419..291B}, which is 
\begin{eqnarray}
\dot{M} &=& \frac{L_{0.1-100}R_{\rm WD}}{GM_{\rm WD}}
 \;=\; \frac{4\pi D^2F_{0.1-100}R_{\rm WD}}{GM_{\rm WD}} \nonumber \\
&=& 
8.4\times10^{16}\;\;\;\mbox{[g s$^{-1}$]}
\label{eq:mdot}
\end{eqnarray}
Inserting Eqs.~(\ref{eq:vb})--(\ref{eq:mdot}) into Eq.~(\ref{eq:x}), we
obtained $h_c = 
7.4 f \times10^{9}$ cm. From the constraint 
$h_c = h_{\rm \Omega} < 0.07~R_{\rm WD}$, we obtain $f < $ 0.007.

\citet{1997MNRAS.291...71H} gave upper limit of $f<0.002$ in the
eclipsing IP XY Arietis which is similar to our result.
According to \citet{2004A&A...426..253R}, $f \simeq 0.01$ based on their
estimation of $h_c = 1.1 \times10^{10}f$ cm from bolometric luminosity
and $\Omega/2\pi<0.5$. This result, however, indicates that the height
of the accretion column is more than 15\% of the WD surface, which
seems unrealistically large for a high accretion rate system like
{\vIZ}.

\section{Conclusion}\label{sec:con}

We have reported results of the Suzaku observation of the typical IP {\vIZ}
in 2007 April. The 5-50 keV X-ray spectra of {\vIZ} can be reproduced
with a multi-temperature optically thin thermal plasma emission model
and its reflection from the WD surface, which suffer photoelectric
absorption represented by a single hydrogen column density, and iron
K$\alpha$ emission lines from the WD surface and the pre-shock
accreting matter as well. The WD spin period is measured to be
$745.7\pm1.1$~s, and the amplitude of the X-ray modulation is deeper in
higher energy bands. This modulation is mainly caused by a rotational
modulation of the hydrogen column density of order $N_{\rm H} \simeq
10^{23}$~cm$^{-2}$, which is twice as large in the spin-minimum phase as in
the spin-maximum phase. This fact is consistent with the accretion curtain model
in which the observer looks down the upper accretion pole in the spin-minimum
phase.

From the spectral analysis, the power index of the temperature
distribution $\alpha$ ($d(EM) \propto T^{\alpha -1}dT$) is found to be
$0.58^{+0.19}_{-0.17}$, which agrees with the value of
0.43 \citep{2005A&A...444..561F} derived based on the full theoretical
calculation of the one-dimensional post-shock accretion flow
\citep{2005A&A...435..191S}. By freezing $\alpha = 0.43$, we have
obtained the maximum temperature of the post-shock plasma to be
$kT_{\rm max} = 37.9^{+5.1}_{-4.6}$~keV, from which we calculated the
mass and the radius of the WD as 
$M_{\rm WD}= 0.82^{+0.05}_{-0.06}~\MO$
and 
$R_{\rm WD} = (6.9\pm 0.4)\times10^8$~cm with the aid of a WD
mass-radius relation. 
We compared our mass value with the other previous works in detail which
show a large mutual inconsistency as large as $0.4~\MO$. We demonstrated
that it is essentially important to take into account the reflection
component appropriately for evaluating the mass of the WD from X-ray
spectroscopy.

We have found a spin modulation of the central energy of the fluorescent
iron K$\alpha$ emission line for the first time in magnetic-CVs
which varies between 6.38~keV and 6.40~keV
with the maximum red-shift at the spin-minimum phase.
More detailed analysis has revealed that the iron line can be decomposed
into a stable 6.4~keV component and another red-shifted component which
manifests itself around the spin-minimum phase at an energy of $\sim$6.3~keV.
The equivalent Doppler velocity of the latter component 
$4.7_{-3.3}^{+2.3}\times10^3$ km sec$^{-1}$ 
and its observed $EW$ of 
$28^{+44}_{-13}$~eV 
are both consistent with an interpretation that it originates from the
pre-shock accreting matter via fluorescence due to irradiation of the
post-shock hot plasma emission. The observed $EW$ of the former stable
component, on the other hand, is $\sim 80$~eV. Considering the ionizing
power of the irradiating thermal spectrum and the observed iron
abundance, we have confirmed that the iron K$\alpha$ line with this
amount of $EW$ can be emitted from the WD surface via fluorescence. The
shock height is evaluated to be small enough compared to the radius of
the WD.

The shock height of the plasma can be evaluated also from the intensity
of the reflected continuum component. The solid angle of the WD viewed
from the post-shock plasma of $\Omega/2\pi = 0.91\pm0.26$
indicates that the shock height is negligibly small
compared with the WD radius ($\Omega/2\pi=1$) as in the case of the
stable component of the iron 6.4~keV line. The upper limit of the shock
height is obtained to be 7~\% of the WD radius ($h_{\rm
K\alpha} < 0.07~R_{\rm WD}$).
Comparing this estimate to the height predicted by a one-dimensional
bremsstrahlung-cooling model of the accretion column,
the fraction area of the WD onto which the accretion really takes place
is limited to be $f <$ 0.007, which is the tightest
constrain on $f$ among the non-eclipsing IPs. This estimate is
comparable to the eclipsing IP XY Ari.

\bigskip

\section{Acknowledgement}

The authors are grateful to all the Suzaku team members 
for their efforts in the production and the maintenance of instruments 
and software, spacecraft operation, and calibrations.
We would like to thank referee for his very useful comments.

\end{document}